\documentclass{iopconfser}
\usepackage{mathptmx}
\usepackage{graphicx}  % needed for figures
\usepackage[mathlines]{lineno}
\usepackage[english]{babel} 
\usepackage{xcolor}
\usepackage{ulem}
\usepackage{dcolumn}   % needed for some tables
\usepackage{bm}        % for math
\usepackage{amssymb}   % for math
\usepackage{amsbsy}   % for math
\usepackage{comment}
\usepackage{multirow}
\usepackage{diagbox}
\usepackage[draft]{todonotes}
\usepackage{mathtools}
\usepackage{newtxtext}
\usepackage{upgreek}
\usepackage[varvw]{newtxmath}
\usepackage{nicefrac}
\usepackage{tikz-feynman}
\tikzfeynmanset{compat=1.1.0}

\begin{document}

\title{Searches for ultra-high energy gamma-ray at the Pierre Auger Observatory and implications on super-heavy dark matter}

\author{Olivier Deligny$^{1}$, for the Pierre Auger Collaboration$^{2}$}

\affil{$^1$Laboratoire de Physique des 2 Infinis Ir\`ene Joliot-Curie (IJCLab)\\
CNRS/IN2P3, Universit\'{e} Paris-Saclay, Orsay, France}
\affil{$^2$Full author list: \texttt{http://www.auger.org/archive/authors\_2024\_06.html}}

\email{deligny@ijclab.in2p3.fr}

\begin{abstract}
The first interactions of photon-induced showers are of electromagnetic nature, and the transfer of energy to the hadron/muon channel is reduced with respect to the bulk of hadron-induced showers. This results in a lower number of secondary muons. Additionally, as the development of photon showers is delayed by the typically small multiplicity of electromagnetic interactions, their maximum of shower development is deeper in the atmosphere than for showers initiated by hadrons. These salient features have enabled searches for photon showers at the Pierre Auger Observatory. They have led to stringent upper limits on ultra-high-energy gamma-ray fluxes over four orders in magnitude in energy. These limits are not only of considerable astrophysical interest, but they also allow us to constrain beyond-standard-physics scenarios. For instance, dark matter particles could be superheavy, provided their lifetime is much longer than the age of the universe. Constraints on specific extensions of the Standard Model of particle physics that meet the lifetime requirement for a superheavy particle will be presented. They include limits on instanton
strength as well as on mixing angle between active and sterile neutrinos.
\end{abstract}

\begin{figure}[t]
\centering
\includegraphics[width=0.49\textwidth]{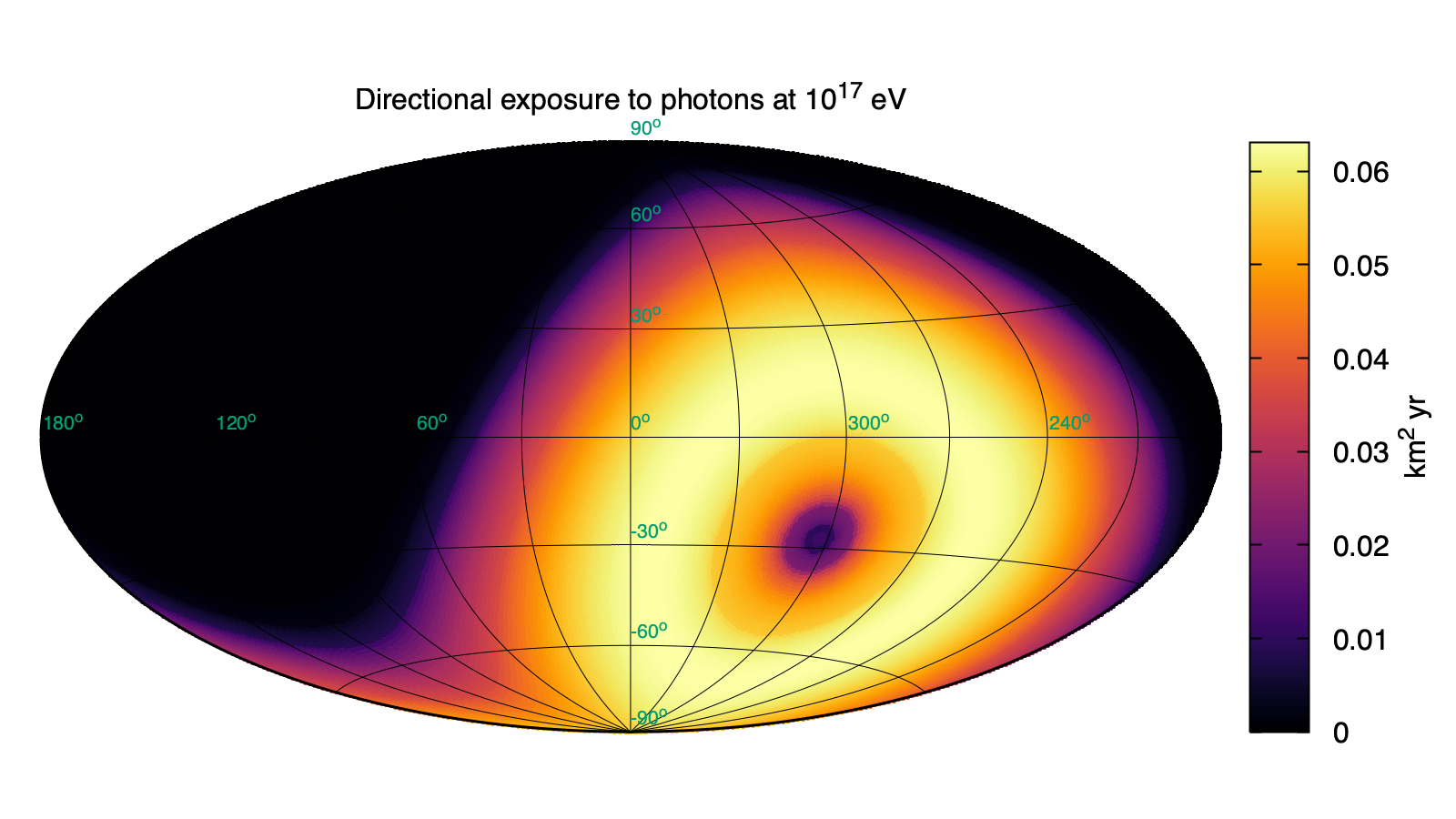}
\includegraphics[width=0.49\textwidth]{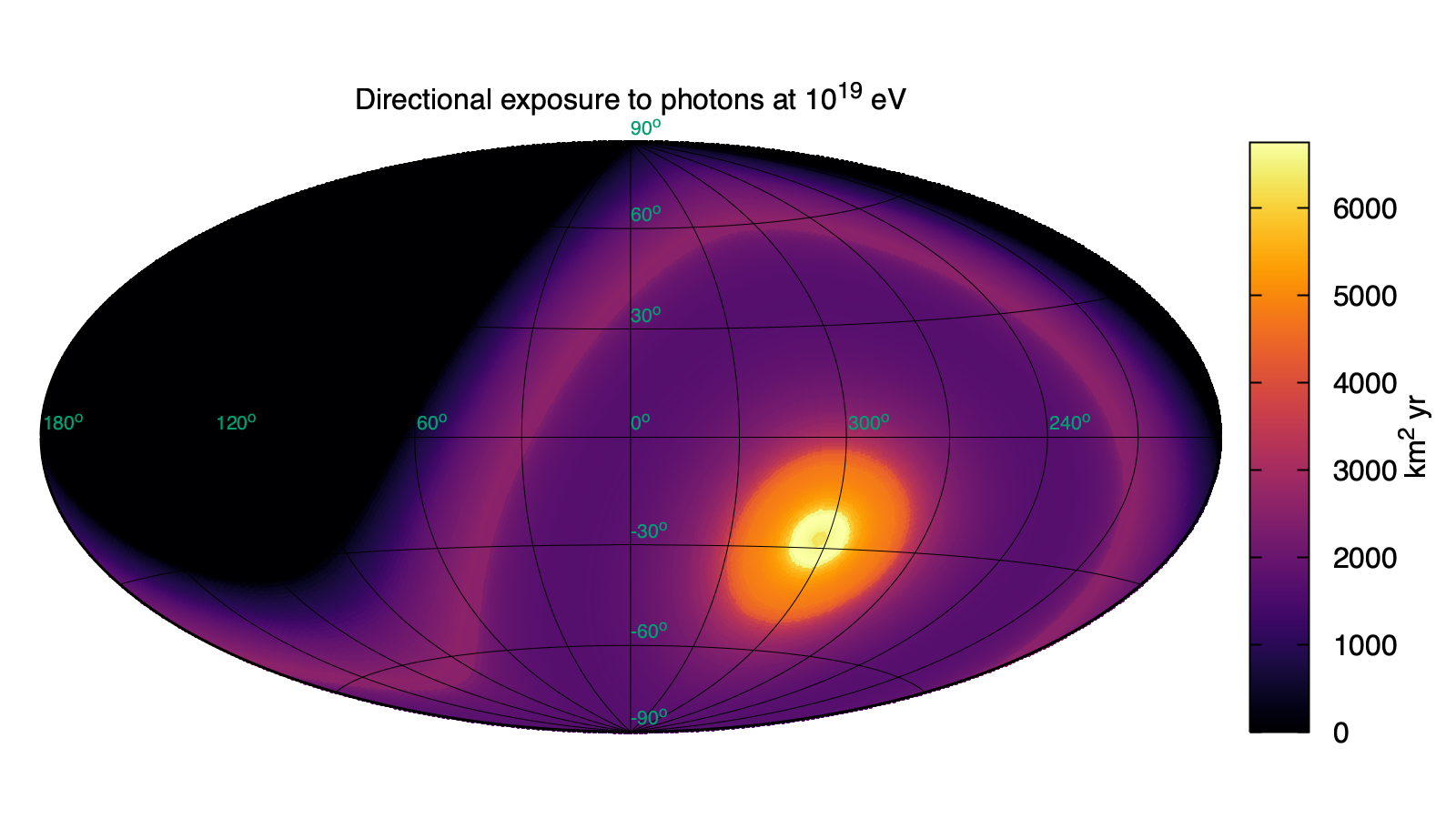}
\caption{Directional exposure of the Pierre Auger Observatory in Galactic coordinates to ultra-high energy gamma-rays at $10^{8}~$GeV (\textit{left}) and $10^{10}~$GeV (\textit{right}). From~\cite{PierreAuger:2023vql}.}
\label{fig:dir_expo_ph}
\end{figure}

\S1~\textit{Searches for ultra-high energy gamma-rays from dark-matter particles.} Compelling evidence for the observation of the decay of super-heavy dark matter particles would be the detection of a flux of gamma-rays with energies in excess of  $10^8~\mathrm{GeV}$, in particular from regions of higher dark-matter density such as the center of our Galaxy. The identification of gamma-ray primaries relies on distinct salient features between the extensive air showers initiated by gamma-rays and those  by the overwhelming background of nuclei: a lower number of secondary muons and deeper $X_\mathrm{max}$ values. Both the ground signal and $X_\mathrm{max}$ can be measured at the Pierre Auger Observatory~\cite{PierreAuger:2015eyc} where a hybrid detection technique is employed for the observation of extensive air showers by combining fluorescence detectors with ground arrays of particle detectors. The combination of the various instruments allows ultra-high-energy gamma-rays to be detected in the energy range above $10^{7.7}~\mathrm{GeV}$ with a directional exposure obtained through the time and area integration of the detection efficiency $\epsilon_\gamma$ and selection efficiency $\kappa_\gamma$ projected onto the direction perpendicular to the arrival
direction,
\begin{equation}
    \label{eqn:expoph}
    \mathcal{E}_\gamma(E,\delta,\alpha)=\int \mathrm{d} \mathbf{x}\int\mathrm{d} t \cos{\theta}\epsilon_\gamma(E,\theta,\mathbf{x},t)\kappa_\gamma(E,\theta).
\end{equation}
In total, four different analyses, differing in the detector used, have been developed to cover the wide energy range probed at the Observatory~\cite{PierreAuger:2023nkh,PierreAuger:2022uwd,PierreAuger:2024ayl,PierreAuger:2022aty}. In particular, two of these analyses benefit from a direct estimate of the depth of shower maximum by the fluorescence detector as one of the discriminating variables. The use of the fluorescence-detector datasets introduces for these analyses an explicit dependence in time for $\epsilon_\gamma$ due to the limited duty cycle on moonless nights that propagates into a small dependence in right ascension $\alpha$ for $\mathcal{E}_\gamma$. In addition to the detection efficiency, the $\kappa_\gamma$ factor accounts for the dependencies of the selection process aimed at separating hadronic-induced showers from gamma-ray-induced ones.  For illustration, we show in Fig.~\ref{fig:dir_expo_ph} an example of directional exposure to photons at $10^{8}~$GeV (left panel) and at $10^{10}~$GeV (right panel). The former is built on the 1.14~km$^2$ array with a separation of detectors of 433~m optimized to study the range of energies around $10^{8}~$GeV, while the latter is the 3,000~km$^2$ one, optimized for higher energies. This explains the large increase of exposure observed at high energies. Overall, no gamma-ray could be firmly identified up to now. The non-observation of candidates (beyond the expected background) allowed the derivation of upper bounds that can constrain various models very effectively, as discussed below.

Due to their attenuation over intergalactic distances, only ultra-high energy gamma-rays emitted in the Milky Way can survive on their way to Earth. The emission rate per unit volume and unit energy $q_\gamma$ from decaying dark-matter particles is shaped by the energy density of dark matter $\rho_\mathrm{DM}$,
\begin{equation}
    \label{eqn:q_gamma}
    q_\gamma(E,\mathbf{x}_\odot+s\mathbf{u}) = \frac{1}{M_X\tau_X}\frac{dN_\gamma}{dE}\rho_\mathrm{DM}(\mathbf{x}_\odot+s\mathbf{u}),
\end{equation}
where $M_X$ is the mass of the dark-matter particle, $\tau_X$ its lifetime, $dN_\gamma/dE$ is the energy spectrum of the gamma-ray decay byproducts, $\mathbf{x}_\odot$ is the position of the Solar system in the Galaxy, and $\mathbf{u}$ is a unit vector on the sphere. The energy density is  normalized to $\rho_\odot=0.45$\,GeV\,cm$^{-3}$~\cite{Jiao:2023aci}. The flux (per steradian) of ultra-high energy gamma-rays produced by the decay of dark-matter particles, $J_{\mathrm{DM},\gamma}(E,\mathbf{u})$, is then obtained by integrating the position-dependent emission rate $q_\gamma$ along the path of the photons in the direction $\mathbf{u}$,
\begin{equation}
    \label{eqn:Jgal}
    J_{\gamma}(E,\mathbf{u})=\frac{1}{4\pi}\int_0^\infty ds~q_\gamma(E,\mathbf{x}_{\odot}+s\mathbf{u}),
\end{equation}
where the $4\pi$ normalization factor accounts for the isotropy of the decay processes. Finally, the expected number of events above a threshold energy follows from
\begin{equation}
    \label{eqn:ngal}
    n_{\gamma}(E)=\int d\mathbf{u}\int_{>E}dE'~\mathcal{E}_\gamma(E',\mathrm{n})J_\gamma(E',\mathbf{u}).
\end{equation}
Assuming that the relic abundance of dark matter is saturated by super-heavy particles, constraints can be inferred in the plane $(\tau_X,M_X)$ by requiring the number of gamma-rays expected from Eq.~\ref{eqn:ngal} to be less than the upper limits on the number of observed events accounting for the expected background. For generic two-body decay channels, the constraints generally lead to lifetime much longer than the age of the universe  for $M_X>10^9$~GeV. \\

\S2~\textit{Lifetime constraints in the perturbative domain.} To comply with the lifetime constraints, some  models postulate the existence of super-weak couplings between the dark and standard-model sectors. The lifetime $\tau_X$ of the particles is then governed by the strength of the couplings $g_{X\Theta}$ (or reduced couplings $\alpha_{X\Theta}=g_{X\Theta}^2/(4\pi)$) and by the mass dimension $n$ of the operator $\Theta$ standing for the standard model fields in the effective interaction~\cite{deVega:2003hh}. Even without knowing the theory behind the decay of the dark-matter particle, generic constraints on $\alpha_{X\Theta}$ and $n$ can be derived~\cite{PierreAuger:2022ubv}. The effective interaction term that couples the field $X$ associated with the heavy particle to the standard-model fields is taken as
\begin{equation}
    \label{eqn:Lint}
    \mathcal{L}_\text{int}=\frac{g_{X\Theta}}{\Lambda^{n-4}}X\Theta,
\end{equation}
where $\Lambda$ is an energy parameter typical of the scale of the new interaction. In the absence of further details about the operator $\Theta$, the matrix element describing the decay transition is considered flat in all kinematic variables so that it behaves as $|\mathcal{M}|^2\sim 4\pi\alpha_{X\Theta}/\Lambda^{2n-4}$. On the basis of dimensional arguments, the lifetime of the particle $X$ is then given as
\begin{equation}
    \label{eqn:tauXTheta}
    \tau_{X\Theta}=\frac{V_n}{4\pi M_X\alpha_{X\Theta}}\left(\frac{\Lambda}{M_X}\right)^{2n-8},
\end{equation}
where $V_n$ is a phase space factor. As a proxy for this factor, we use the expression derived for $n-1$ particles in the final state~\cite{Kleiss:1985gy},
\begin{equation}
    \label{eqn:Vn}
    V_n=\left(\frac{2}{\pi}\right)^{n-1}\Gamma(n-1)\Gamma(n-2),
\end{equation}
with $\Gamma(x)$ the Euler gamma function.
\begin{figure}
\centering
\includegraphics[width=0.5\columnwidth]{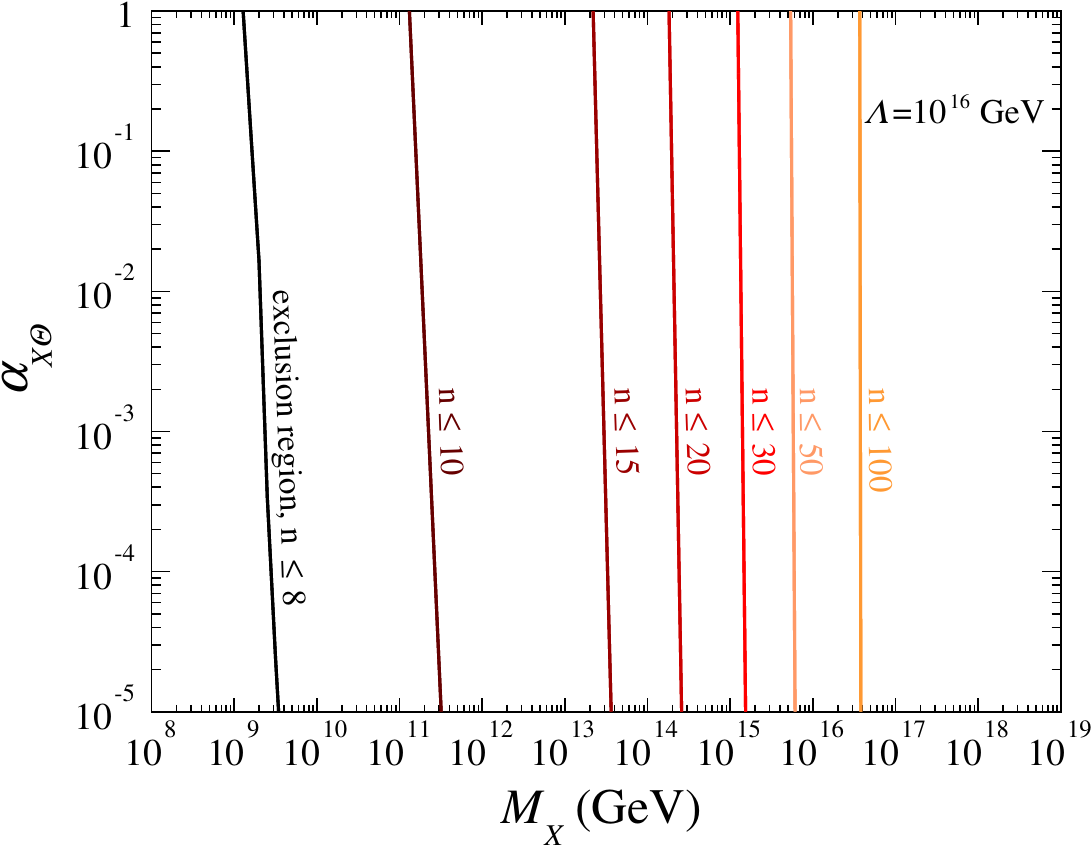}
\caption{Exclusion regions in the plane $(\alpha_{X\Theta},M_X)$ for several values of mass dimension $n$ of operators responsible for the perturbative decay of the super-heavy particle, and for an energy scale of new physics $\Lambda=10^{16}$\,GeV. From~\cite{PierreAuger:2022ubv}.} 
\label{fig:alphaX-mass-pert}
\end{figure}
It is apparent that the coupling constant $\alpha_{X\Theta}$ and the dimension $n$ have to take specific values for super-heavy particles to be stable enough.  In practice, for a specific upper limit on $n_\gamma(E)$ at one energy threshold $E$, a lower limit of the $\tau_{X\Theta}$ parameter is derived for each value of mass $M_X$. The lower limit on $\tau_{X\Theta}$ is subsequently transformed into an upper limit on $\alpha_{X\Theta}$ by means of Eq.~\eqref{eqn:tauXTheta}. By repeating the procedure for each upper limit on $n_\gamma(E)$, a set of curves is obtained, reflecting the sensitivity of a specific energy threshold to some range of mass. The union of the excluded regions finally provides the constraints in the plane $(\alpha_{X\Theta},M_X)$. In this manner we obtain the contour lines shown in Fig.~\ref{fig:alphaX-mass-pert} for several values of $n$ and for an emblematic choice of GUT $\Lambda$ value. The scale chosen for $\alpha_{X\Theta}$ ranges from 1 down to $10^{-5}$. It is observed that for the limits on photon fluxes to be satisfied, the mass of the super-heavy particle cannot exceed ${\gtrsim}10^9$\,GeV (${\gtrsim}10^{11}$\,GeV) for operators of dimension equal to or larger than $n=8$ ($n=10$), while larger masses require an increase in $n$. To approach the large masses while keeping operators of dimension relatively low, ``astronomically-small'' coupling constants should be at work. The same conclusions hold for other choices of $\Lambda$. All in all, for perturbative processes to be responsible for the decay of super-heavy dark matter particles requires quite ``unnatural'' fine-tuning.   \\

\S3~\textit{Constraints on instanton-induced decays.} Stability of super-heavy dark matter particles is consequently calling for a new quantum number to be conserved in the dark sector so as to protect the particles from decaying. Nevertheless, even stable particles in the perturbative domain will in general eventually decay due to non-perturbative effects (instantons) in non-abelian gauge theories. Instanton-induced decay can thus make observable a dark sector that would otherwise be totally hidden by the conservation of a quantum number~\cite{Kuzmin:1997jua}. Assuming quarks and leptons carry this quantum number and so contribute to anomaly relationships with contributions from the dark sector, they will be secondary products in the decays of dark matter together with the lightest hidden fermion. The lifetime of the decaying particle is mainly driven by the instanton suppression factor that leads to~\cite{tHooft:1976rip} 
\begin{equation}\label{eqn:tauX}
    \tau_X \simeq M_X^{-1}\exp{\left(4\pi/\alpha_X\right)},
\end{equation} 
with $\alpha_X$ the reduced coupling constant of the hidden gauge interaction. 

\begin{figure}
\centering
\includegraphics[width=0.6\columnwidth]{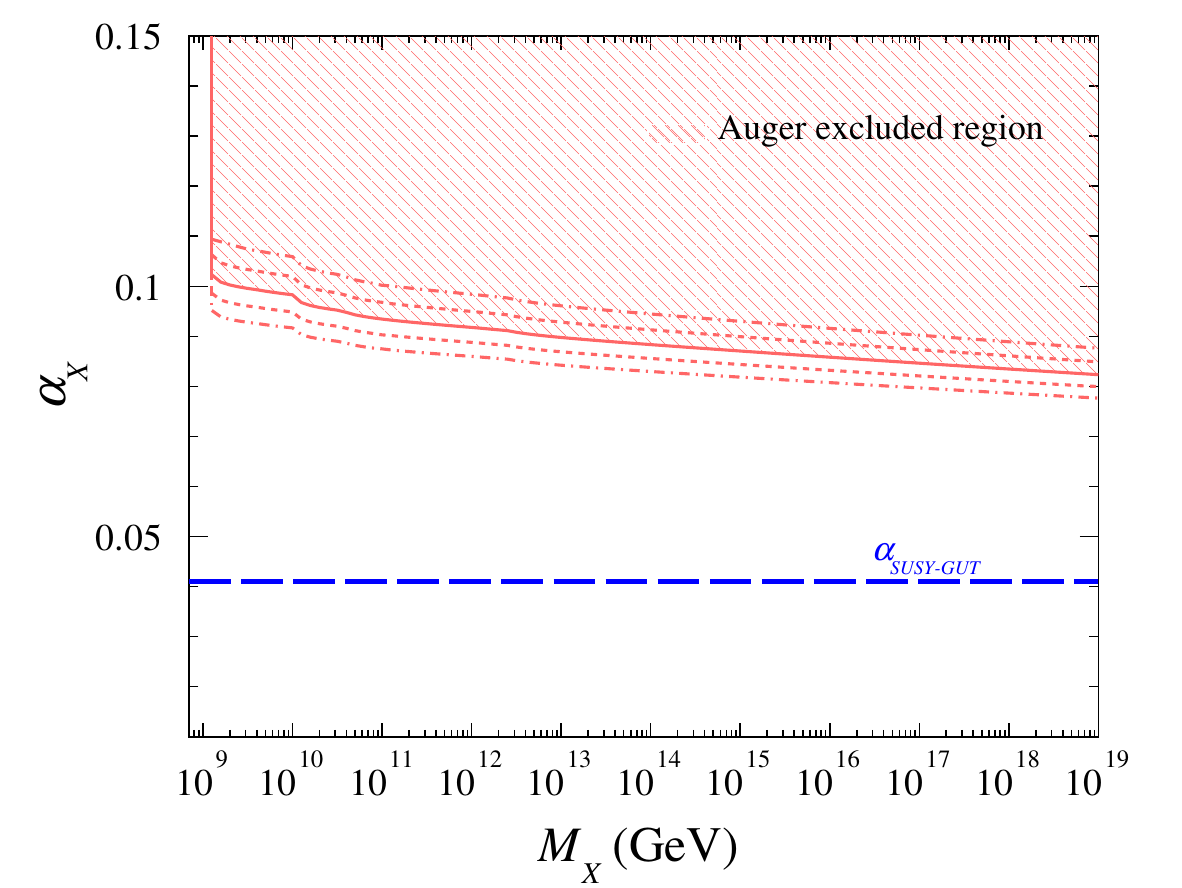}
\caption{Upper limits at 95\% C.L. on the  coupling constant $\alpha_X$ of a hidden gauge interaction as a function of the mass $M_X$ of a dark matter particle $X$ (assumed to compose 100\% of the observed dark matter abundance) decaying into a dozen of $q \bar q$ pairs. From~\cite{PierreAuger:2022jyk}.}
\label{fig:alphaX-mass}
\end{figure}

Quite independently of the hidden gauge interaction, the exact content in instanton-induced decays of quarks and leptons, which will eventually produce hadrons decaying into photons and neutrinos, obeys selection rules that involve very large multiplicities. Assuming ten pairs  of quarks and leptons in the final state, the energy spectrum  $dN_\gamma(E,M_X)/dE$ then follows from fragmentation effects~\cite{Aloisio:2003xj,Sarkar:2001se,Barbot:2002gt,Bauer:2020jay}. 

Eq.~\eqref{eqn:tauX} provides us with a relationship connecting the lifetime $\tau_X$ to the coupling constant $\alpha_X$. In the same way as in the perturbative case above, upper limits on $\alpha_X$ can be obtained. They are shown as the shaded red area in Fig.~\ref{fig:alphaX-mass}. The coupling should be less than $\simeq 0.09$ for a wide range of masses. Numerical factors in front of the exponential could however arise in Eq.~(\ref{eqn:tauX}) depending on the underlying model for the hidden gauge sector. Explicit constructions of the dark sector are, however, well beyond the scope of this contribution. Although the limits presented in Fig.~\ref{fig:alphaX-mass} are hardly destabilized due to the exponential dependence in $\alpha_X^{-1}$, we note that model-dependent numerical factors in front of the exponential such as $10^{\pm k}$ lead to a shift of $\pm 0.0013k$ in $\alpha_X$ limits. We show in dotted and dashed lines the bounds for $k=2$ and $k=4$, respectively. These factors are by far the dominant systematic uncertainties.\\

\S4~\textit{Constraints on dark matter coupled to sterile neutrinos.} A superheavy particle that is metastable can also result from the coupling between a pseudo-scalar particle with sterile neutrinos embedded in an extended-seesaw framework~\cite{Dudas:2020sbq}. In this beyond-standard-model extension,  the dark-matter particle $X$ interacts only with sub-eV and superheavy ($10^{12-14}~$GeV) sterile neutrinos, of masses $m_N$ and $M_N$ respectively, via Yukawa couplings $y_m$ and $y_M$. In the mass-eigenstate basis, neutrinos $\nu_1$ and $\nu_2$ are then quasi-sterile or quasi-active respectively, depending on the mixture of active and sterile neutrinos governed by a small mixing angle $\theta_m\simeq \sqrt{2} y_m v/m_\nu$, with $v$ the electroweak scale (mass dimension) and $m_\nu$ the mass of the known neutrinos. To leading order in $y_m$, quasi-active neutrinos are produced from quasi-sterile ones subsequent to the decay of $X$. Consequently, the coupling $y_m$ controls the dominant decay channels and allows for trading a factor $(M_X/M_\mathrm{P})^2$ (with $M_\mathrm{P}$ the Planck mass) for a $(m_\nu\theta_m/v)^2$ one in the decay width of $X$. This trading enables the reduction of the width by a factor $\sim 10^{-25}\theta_m^2$ for a benchmark value $M_X=10^9~$GeV, leading to the required lifetimes well beyond the age of the universe. 

\begin{figure}[t]
\centering
\includegraphics[width=0.6\textwidth]{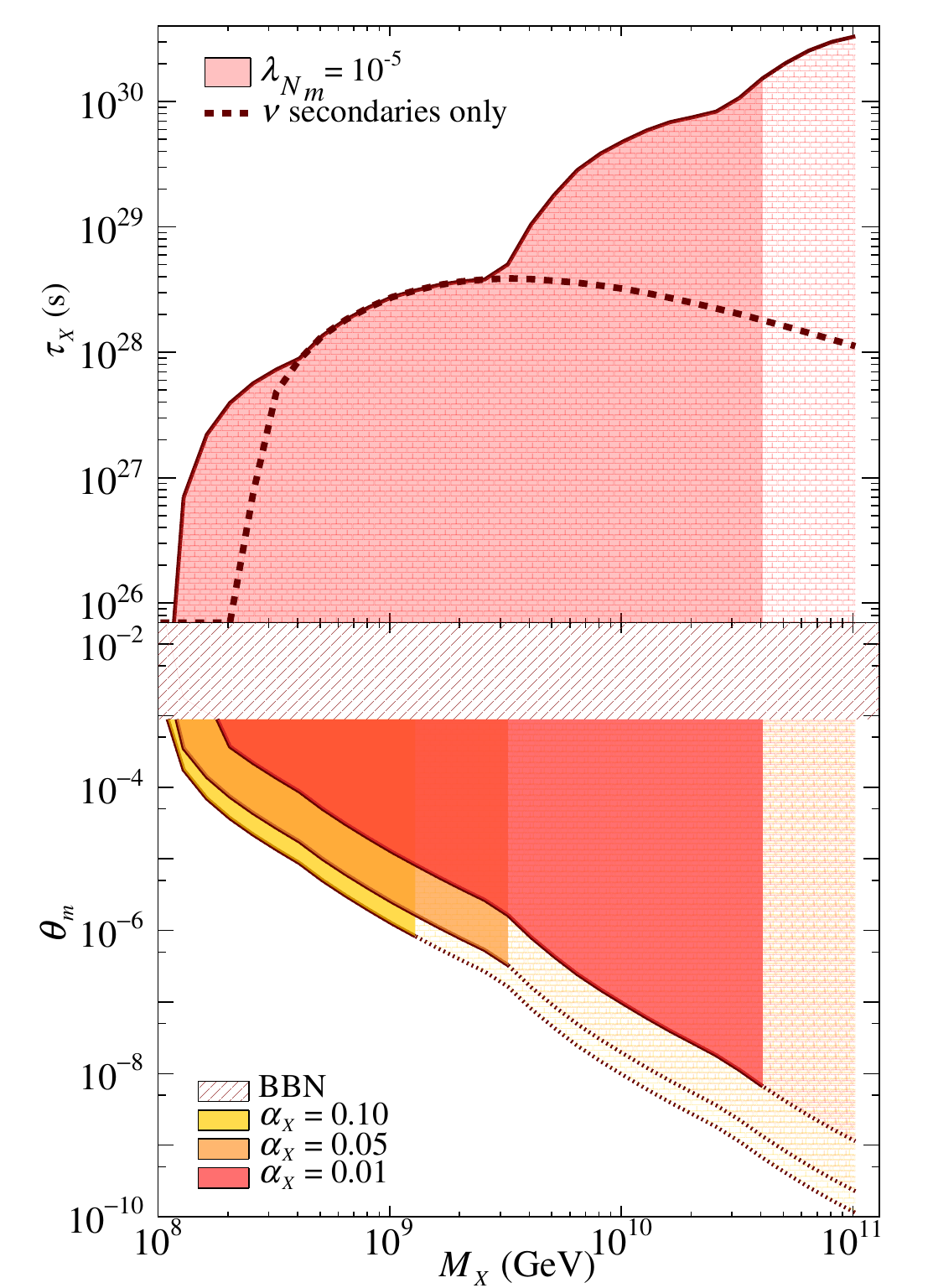}
\caption{\textit{Top}: Constraints on $\tau_X$ as a function of $M_X$ for a value of the couplings of sterile neutrinos with the inflationary sector $\lambda_{N_m}=10^{-5}$. The dotted line illustrates the constraints stemming from neutrino secondaries alone. \textit{Bottom}: Constraints on $\theta_m$ as a function of $M_X$ for three different values of the coupling constant $\alpha_{X}$, still for $\lambda_{N_m}=10^{-5}$. The hatched-red region $\theta_m\geq 9{\times}10^{-4}$ is excluded from the constraint on $\Delta N_\mathrm{eff}$ (see text). From~\cite{PierreAuger:2023vql}.}
\label{fig:result}
\end{figure}
To leading order in $\theta_m$, the total width $\Gamma^X$ is dominated by three-body channels stemming from the interaction between active/sterile neutrinos and the Higgs isodoublet with Yukawa coupling $y_m\simeq \sqrt{2}\theta_mm_2/v$. The channel $X\rightarrow h\nu_1\nu_2$, diagrammatically represented as
\begin{equation}
\begin{tikzpicture} \begin{feynman}
   \vertex(a) {\(X\)};
   \vertex[right=of a] (b);
   \vertex[above right=of b] (f1){\(\nu_{1}\)};
   \vertex[below right=of b] (c);
   \vertex[above right=of c] (f2){\(h\)};
   \vertex[below right=of c] (f3){\(\nu_{2}\)};
   \diagram* {
    (a)-- [scalar] (b)-- [anti fermion] (f1),
    (b)-- [fermion,edge label'=\(\nu_{1}\)] (c),
    (c)-- [scalar] (f2),
    (c)-- [fermion] (f3),
   };
   \node at (3.6,-1.1) {\(\theta_m m_2/v,\)};
   %\node at (3.2,-0.7) {\((\alpha_{X\nu_{s}}/M_{P})\partial_\mu\gamma^\mu\gamma^5\)};
\end{feynman}\end{tikzpicture}
\label{eqn:3body-Xhnunu}
\end{equation}
gives the most important contribution to the width~\cite{Dudas:2020sbq}:
\begin{equation}
\Gamma_{h\nu_1\nu_2}^X=\frac{\alpha_X^2\theta_m^2}{192\pi^3}\left(\frac{M_X}{M_\mathrm{P}}\right)^2\left(\frac{m_2}{v}\right)^2 M_X. 
\label{eqn:width-hnunu}
\end{equation}
As announced, a factor $(M_X/M_\mathrm{P})^2$ expected from dimensional arguments is indeed traded for a factor $(m_2/v)^2$. 
Constraints in the planes $(\tau_X,M_X)$ and $(\theta_m,M_X)$ can be derived from the non-observation of ultra-high energy gamma-rays (and neutrinos) at the Observatory. First, 90\% C.L. lower limits on the lifetime $\tau_X$ are obtained as previously by setting, for a specific value of $M_X$, $n_\gamma(E)$ or $n_\nu(E)$ to the 90\% C.L. upper-limit numbers corresponding to the number of background-event candidates in the absence of signal. Subsequently, a scan in $M_X$ is carried out. It leads to a curve in the plane $(\tau_{X},M_X)$ that pertains to the energy threshold $E$ considered. By repeating the procedure for several thresholds, a set of curves is obtained, reflecting the sensitivity of a specific energy threshold to some range of mass $M_X$. The union of the excluded regions finally provides the constraints in the $(\tau_{X},M_X)$ plane. Results are shown in Fig.~\ref{fig:result} (top panel); lifetimes within the cross-hatched region are excluded. The region in full red pertains to a particular value of a Yukawa coupling $\lambda_{N_m}=10^{-5}$, the meaning of which will be explained below. To illustrate the contribution from each secondary at our disposal, we show as the dotted the contribution to the constraints stemming from neutrinos alone; an analysis of the IceCube exposure dedicated to the benchmark-scenario decay channels would likely provide better sensitivity for exploring masses $M_X\lesssim 10^{8.5}~$GeV. The lower limit on $\tau_X$ is then transformed into an upper limit on $\theta_m$ using the expressions of the total width of the particle $X$. Results are shown in the bottom panel for separate values of $\alpha_X$: color-coded regions pertain to $\lambda_{N_m}=10^{-5}$ while their extension (in cross-hatched) would require smaller values of  $\lambda_{N_m}$. Systematic uncertainties on $\theta_m$ constraints amount overall to $\simeq \pm 15\%$; they are dominated by those on the neutrino exposure~\cite{PierreAuger:2019ens}. The restricted ranges of $M_X$ for different $\alpha_X$ values come from the requirements not to overclose the universe with dark matter, while the exclusion hatched band comes from not altering the expansion history of the universe with the presence of ultra-light species such as sterile neutrinos $N_m$. We now briefly explain how these constraints are obtained. \\

\begin{figure}
\centering
\includegraphics[width=0.49\columnwidth]{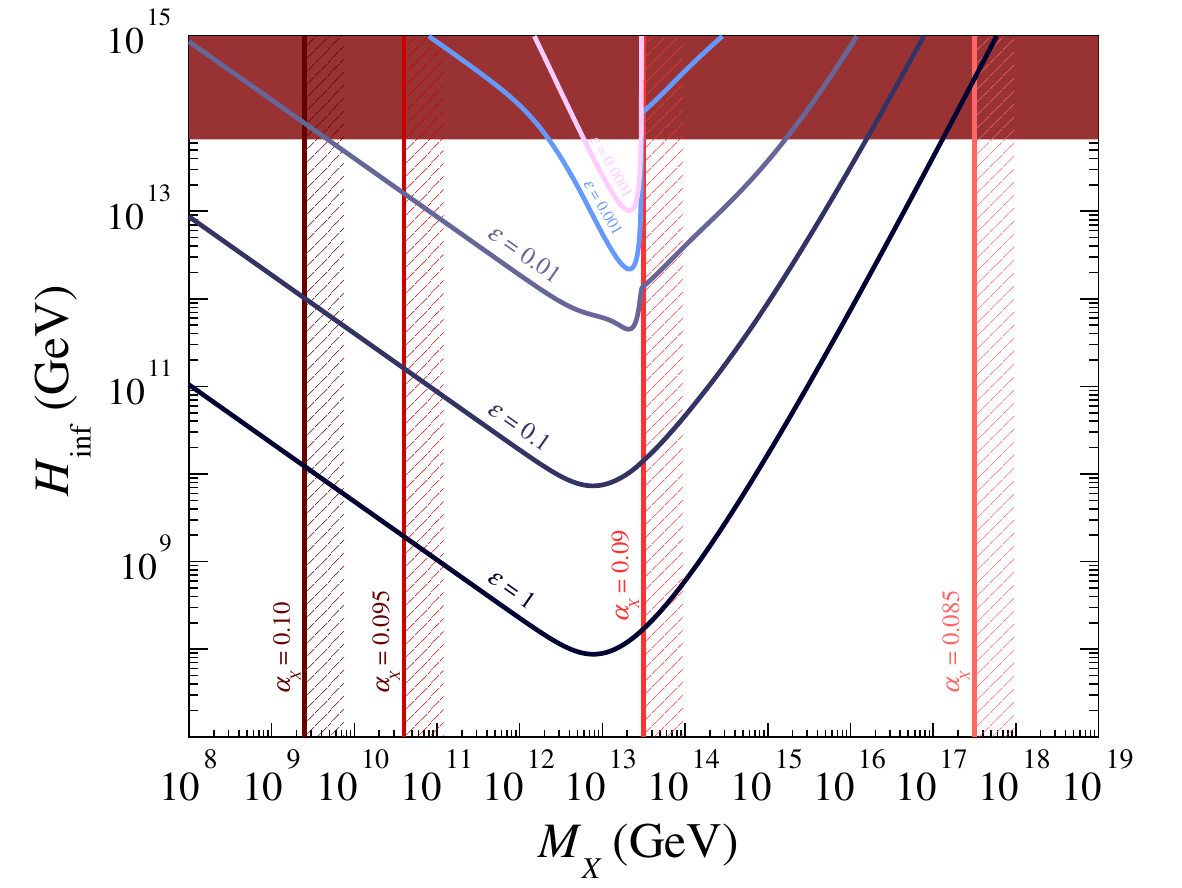}
\includegraphics[width=0.49\columnwidth]{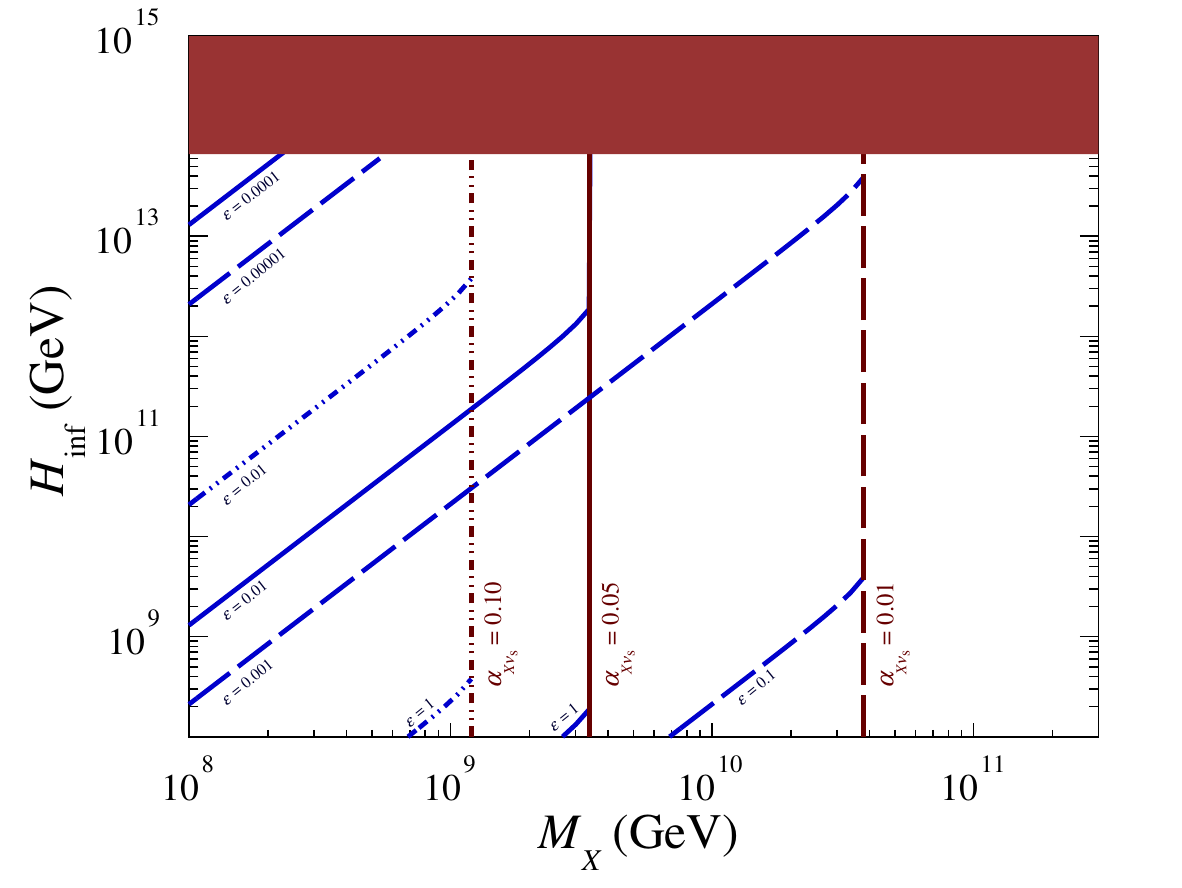}
\caption{Left: Constraints in the $(H_\mathrm{inf},M_X)$ plane, where viable values are delineated by the blue lines for different values of reheating efficiency $\epsilon$. Additional constraints from the non-observation of instanton-induced decay of super-heavy particles allow for excluding the mass ranges in the red-shaded regions, for the specified value of the dark-sector gauge coupling. From~\cite{PierreAuger:2022jyk}. Right: Same, adding the possibility of a radiative production of dark matter in the inflaton decay.}
\label{fig:Hinf}
\end{figure}

\S5~\textit{Cosmological constraints.} Gravitational interaction alone may have been sufficient to produce the right amount of super-heavy dark matter particles at the end of the inflation era for a wide range of high masses, up to $M_\mathrm{GUT}$, accounting for the production by annihilation of standard model particles (SM)~\cite{Garny:2015sjg} or of inflaton particles ($\phi$ hereafter)~\cite{Mambrini:2021zpp} through the exchange of a graviton. In this scenario, the relic abundance of super-heavy dark matter  particles can be estimated from the quite involved reheating dynamics~\cite{Chung:1998rq,Giudice:2000ex}. The time evolution of the $X$-particle density $n_X$ reads as
\begin{equation}
\label{eqn:nX}
\frac{dn_X(t)}{dt}+3H(t)n_X(t)\simeq \sum_i \overline{n}_i^2(t)\gamma_i,%\langle\sigma v\rangle ~\overline{n}_X^2(t),
\end{equation}
where the sum in the right hand side stands for the contributions from the standard model~\cite{Garny:2015sjg} and inflationary~\cite{Mambrini:2021zpp} sectors to produce fermions. Introducing the dimensionless abundance $Y_X=n_Xa^3/T_\mathrm{rh}^3$ to absorb the expansion of the universe, with $T_\mathrm{rh}$ the reheating temperature, and using $aH(a)dt=da$ from the definition of the Hubble parameter (with $a$ the scale factor), Eq.~\ref{eqn:nX} becomes
\begin{equation}
\label{eqn:YX}
\frac{dY_X(a)}{da}\simeq\frac{a^2}{T_\mathrm{rh}^3H(a)}\sum_i \overline{n}_i^2(a)\gamma_i,
\end{equation}
which, using the dynamics of the expansion rate during reheating, yields the present-day dimensionless abundance $Y_{X,0}$ assuming $Y_{X,\mathrm{inf}}=0$. The present-day relic abundance, $\Omega_\mathrm{CDM}$, can then be related to $M_X$, $H_\mathrm{inf}$, and $\epsilon=T_\mathrm{rh}/(0.25\sqrt{M_\mathrm{P}H_\mathrm{inf}})$ through~\cite{Garny:2015sjg} $\Omega_\mathrm{CDM}h^2=9.2~10^{24}\epsilon^4M_XY_{X,0}/M_\mathrm{P}$. As a result, one interesting viable possibility in the $(H_\mathrm{inf},M_X)$ parameter space is that $X$ particles with masses as large as the GUT energy scale could be sufficiently abundant to match the dark matter relic density, provided that the inflationary energy scale is high ($H_\mathrm{inf} \sim 10^{13}~\mathrm{GeV}$) and the reheating efficiency is high (so that reheating is quasi-instantaneous). This rules out values of the dark-sector gauge coupling greater than ${\simeq}0.085$, as observed in the left panel of Fig.~\ref{fig:Hinf}. The mass values could however be smaller if the reheating temperature is not that high. In general, for high efficiencies $\epsilon$ (corresponding to short duration of the reheating era), the $\mathrm{SM} + \mathrm{SM} \to X+X$ reaction allows for a wide range of $M_X$ values. For efficiencies below $\simeq 0.01$, the $\phi + \phi \to X+X$ reaction allows for solutions in a narrower range of the $(H_\mathrm{inf},M_X)$ plane close to $M_X=10^{13}~\mathrm{GeV}$, with in particular $M_X \leq M_\phi$ as a result of the kinematic suppression in the corresponding rate $\gamma_i$~\cite{Mambrini:2021zpp}.

Similar constraints can be drawn for the model invoking couplings of the $X$ particle with sterile neutrinos. In addition to its couplings to the dark-matter sector and to the standard-model  one through the Higgs isodoublet, the sterile neutrino $N_m$ is also coupled to an inflationary sector in the benchmark~\cite{Dudas:2020sbq}. This coupling, governed by a unique Yukawa parameter $\lambda_{N_m}$ for every $\nu_1$ neutrinos, yields to a ``radiative'' production of $X$ via a diagram similar to that depicted in Eq.~\ref{eqn:3body-Xhnunu} (substituting $X$ by the inflaton $\Phi$ in the initial state, and $h$ and $\nu_2$ by $X$ and $\nu_1$ in the final states). Such a mechanism leads to a direct production of dark matter during the reheating period that can be sufficient, in general, to match the right amount of dark matter observed today~\cite{Kaneta:2019zgw}. In the  benchmark~\cite{Dudas:2020sbq}, values for $\lambda_{N_m}$ are then required to range preferentially around $10^{-5}$.  To infer the dark matter density $n_X$ produced mainly during the reheating epoch, we also consider the minimal setup of gravitational production of $X$ particles through the annihilation of standard-model (inflaton) particles as in~\cite{Garny:2015sjg} (as in~\cite{Mambrini:2021zpp}). In these conditions, $X$ particles can be produced as long as the collision rate of particles is larger than the expansion rate $H$ and/or as long as the inflaton field oscillates. By contrast, $n_X$ is prohibitively low to allow any thermal equilibrium for dark matter. The collision term in the Boltzmann equation is then approximated as a source term only. Overall, the Boltzmann equation reads as
\begin{eqnarray}
   \label{eqn:boltzmann}
   \frac{dn_X(t)}{dt}+3H(t)n_X(t)=\sum_i \overline{n}^2_i(t)\gamma_i+\overline{n}_\phi(t)\Gamma_{X\nu_1\nu_1}.
\end{eqnarray}
As a result, viable couples of values for $(H_\text{inf},M_X)$ scale as $H_\text{inf} \propto M_X^2$ up to a maximum value for $M_X$, which depend on $\epsilon$ and $\alpha_X$ -- see right panel of Fig.~\ref{fig:Hinf}. This scaling is a consequence of the domination of the radiative-production process over the gravitational one as long as the allowed values of $H_\text{inf}$ are too small for a given $M_X$ value to generate significant particle production by gravitational interactions. For larger masses, the contribution from gravitational interactions added to the radiative production of $X$ leads to an overproduction of dark matter that overcloses the universe, and there is thus no longer solution. This explains why the color-coded regions extend up to some maximum values of $M_X$ in Fig.~\ref{fig:result}, for a benchmark value of $\lambda_{N_m}=10^{-5}$. To the right of the regions shown in cross-hatched, $\lambda_{N_m}$ would need to be smaller. \\

\S6~\textit{Conclusions.} We have shown that the data of the Pierre Auger Observatory provide stringent constraints on decaying super-heavy particles that would constitute dark matter. These constraints allow us to put a limit on instanton strength or on 
the angle $\theta_m$ mixing sub-eV sterile and active neutrinos in the context of an extension to the SM that couples the sterile neutrinos to a superheavy DM candidate. Other physics providing mechanisms to stabilize super-heavy particles can be explored, such as supersymmetry broken at high scale with a tiny $R$-parity violation. This will be the subject of a future study.  

\newpage

\bibliographystyle{iopart-num}
\bibliography{bibliography}

\end{document}